# Observation of photoelectric nonvolatile memory and oscillations in VO$_2$ at room temperature

Youngho Jung[1,2]*, Junho Jeong[1,2]*, Zhongnan Qu[2], Bin Cui[1], Ankita Khanda[2], Stuart S. P. Parkin[1], and Joyce K. S. Poon[1,2]

**Affiliations:**

[1]Max Planck Institute of Microstructure Physics, Weinberg 2, Halle, 06120, Germany.

[2]Department of Electrical and Computer Engineering, University of Toronto, 10 King's College Rd., Toronto, Ontario M5S 3G4, Canada.

*These authors contributed equally to this work.

**Abstract:** Vanadium dioxide (VO$_2$) is a phase change material that can reversibly change between high and low resistivity states through electronic and structural phase transitions. Thus far, VO$_2$ memory devices have essentially been volatile at room temperature, and nonvolatile memory has required non-ambient surroundings (e.g., elevated temperatures, electrolytes) and long write times. Here, we report the first observation of optically addressable nonvolatile memory in VO$_2$ at room temperature with a readout by voltage oscillations. The read and write times had to be kept shorter than about 150 μs. The writing of the memory and onset of the voltage oscillations had a minimum optical power threshold. This discovery demonstrates the potential of VO$_2$ for new computing devices and architectures, such as artificial neurons and oscillatory neural networks.

**INTRODUCTION**

Phase-change materials (PCMs) are important candidates for next generation computing devices and systems, including artificial neurons and neuromorphic computing (*1-10*) that aim to improve power consumption and speed in data transfer (*2-7*). A PCM gaining much interest for such applications is the transition metal oxide, vanadium dioxide (VO$_2$), which can reversibly change between its insulator and metal states by thermal (*11*), electrical (*12, 13*), optical (*14*), or mechanical stimuli (*15*) or by ionic liquid gating (*16, 17*). The phase transition has an hysteresis that is utilized for memory (*15-20*), and the negative differential resistance (NDR) regime of the $VO_2$ leads to voltage or current oscillations (*12, 21, 22*).

Thus far, the $VO_2$ insulator-metal transition (IMT) due to a structural phase transition (SPT) has been volatile at room temperature, making a $VO_2$ memory cell power-inefficient. Nonvolatile phase-change has been observed in $VO_2$ on a piezoelectric substrate, but the non-volatility is due to remnant strains in the piezoelectric material rather than the $VO_2$ itself (*15*). Tuning the oxygen vacancy (*16*) or hydrogenation (*17*) of $VO_2$, using a gated electrolyte, can lead to nonvolatile, reversible phase transitions but, to date, is very slow (requiring minutes). Other observations of nonvolatile $VO_2$ memories required an external bias, in the form of a raised temperature or applied voltage, to maintain their memory states (*18-20*). Here, we report the observation of nonvolatile $VO_2$ memory at room temperature with sub-millisecond writing and reading times. The memory was addressed optically, and the written memory could be read out as voltage oscillations. We term this device a $VO_2$ "photo-electric memory-oscillator" (PEMO). It did not require a persistent external bias to maintain the memory state and operated at room temperature in the ambient environment.

**RESULTS**

**Overview**

We begin with an overview of the device operation and the observed phenomenon. The $VO_2$ PEMO consisted of two electrical terminals patterned on an etched $VO_2$ microwire (Fig. 1A). To prepare the device for memory writing, it was first held at a constant current bias in the insulating-monoclinic phase ($M_1$) near the Mott transition. To write the memory, an excitation, which was a weak optical pulse at a wavelength of 1550 nm, was applied. The excitation generated excess carriers to drive the $VO_2$ into the intermediate oscillating state (O) between the first Mott transition and the second Peierls SPT. In this regime, the device exhibited NDR (see Fig. 1A) and thus voltage oscillations between the $M_1$ and metallic-rutile (R) phases of $VO_2$ (*12*). After the writing with an optical pulse, the current bias was turned off within a fall-time of <150 μs. Interestingly, we found that when the memory-writing bias current was applied again with a sufficiently short rise-time of <180 μs, contrary to the expectation of the $VO_2$ returning to the $M_1$ phase, voltage oscillations could be read out again even though no current or voltage was applied between the writing and reading steps. The device could be reset by applying a current less than the hysteresis of the Mott transition.

**Experimental Setup and Material Properties**

Next, we discuss the details of the experimental investigation. Figure 1A shows the schematic of the device along with the experimental setup (see fig. S1 for more details). The microwire $VO_2$ device was defined using conventional microfabrication processes in a 360 nm thick $VO_2$ film on a titanium dioxide ((001) $TiO_2$) substrate formed by pulsed laser deposition. Figure 1B shows a scanning electron micrograph of a device. The $VO_2$ film on (001) $TiO_2$ substrate was strained along the *c*-axis by –1.2% due to the lattice mismatch (*16*), which resulted in a phase transition temperature ($T_C$) of ~41.5 °C (314.65 K) as shown in Fig. 1C. Across the phase transition, the $VO_2$ resistivity changed by about 3 orders of magnitude. Room-temperature X-ray diffraction (XRD) measurement (inset of Fig. 1C) confirms the composition of the material to be $VO_2$ (*23*), and an atomic force microscopy shows that the film has a root-mean-square roughness of ~0.55 nm (fig. S2A).

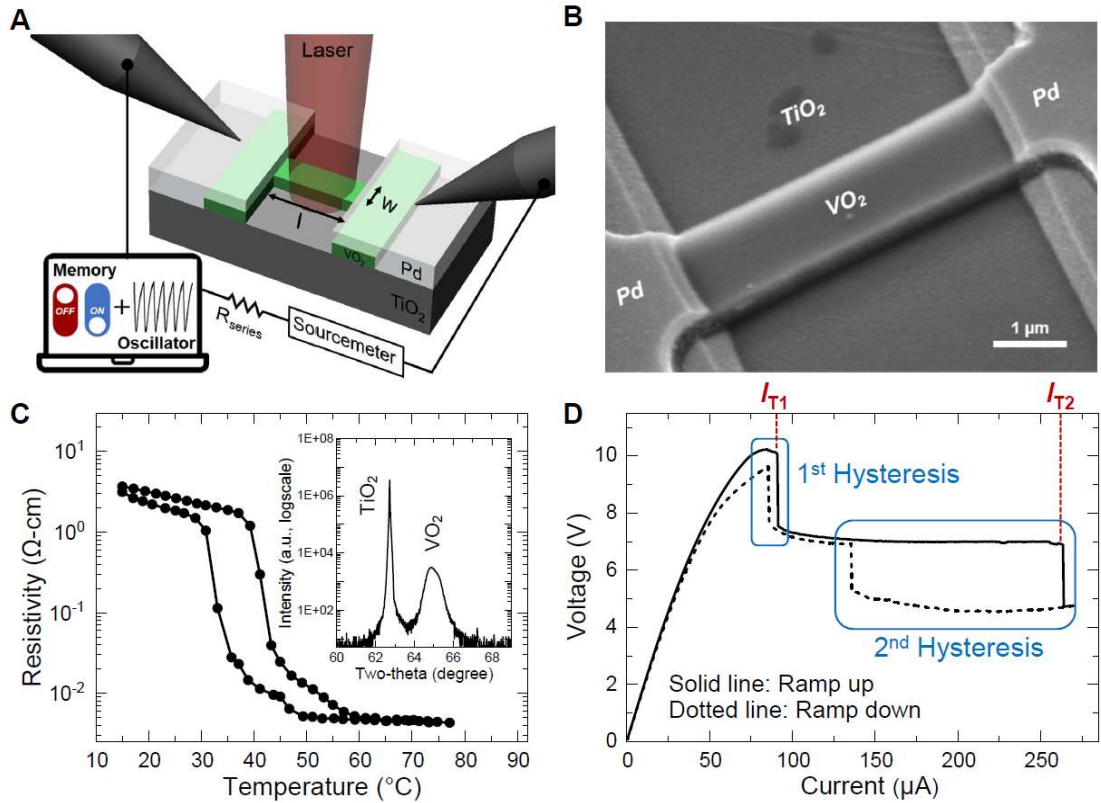

**Fig. 1. Experimental preparation and characterization for VO$_2$ photo-electric memory-oscillator (PEMO).** (**A**) 3D schematic of the device and experimental setup. (**B**) A scanning electron micrograph of a representative VO$_2$ microwire with Pd contacts. (**C**) Measured resistance of a VO$_2$ device as a function of stage temperature. Inset, x-ray power diffraction (XRD) spectrum of a VO$_2$ film on a TiO$_2$(001) substrate. (**D**) Typical example of VI measurement at room temperature, showing two abrupt transitions. The 1$^{st}$ and 2$^{nd}$ transitions corresponds to Mott transition and SPT, respectively. The phase transition currents, $I_{T1}$ and $I_{T2}$, are labelled.

**Device Voltage-Current Characteristics**

The memory measurements were conducted at room temperature and in the ambient environment. The DC voltage-current (VI) plot for a device with VO$_2$ wire dimensions $w$ = ~1.7 μm and $l$ = ~4.7 μm is shown in Fig. 1D. We ramped the current between 0 and 260 μA in 500 nA increments with a resistor connected in series ($R_{series}$ = 5 kΩ). As can be seen in the DC VI plot, when the current was increased (solid line), the voltage across the device dropped from ~10 V to ~4.7 V through the 1st ($I_{T1}$ ~ 91 µA) and 2nd ($I_{T2}$ ~264 µA) transitions, which constituted the full phase transition from the $M_1$ to R phases including SPT. When the current was decreased (dashed line), the reverse metal-insulator transition occurred at different currents; thus, both the 1st and 2nd transitions were hysteretic, with a smaller hysteresis width for the 1st transition and the larger hysteresis width at the 2nd transition due to the predominantly thermal-driven SPT.

**Nonvolatile Memory and Voltage Oscillations**

To investigate the nonvolatile memory, we used a combination of an electrical current bias and an optical writing pulse in the 1st transition. Figure 2A shows the DC VI plot of the 1st transition of the device corresponding to Fig. 1D at a temperature of 23 °C. The hysteresis loop was between $I_{H1}$~82.4 µA and $I_{H2}$~87.6 µA, obtained by repeating the measurements 10 times. Before writing to the memory, we held the VO$_2$ microwire at a current bias of $I_B$~85 µA (denoted by Point 1 in Fig. 2A) for ~3 seconds. To write to the device, we turned on a ~180 ms-long laser pulse at a wavelength of 1550 nm, incident power of ~470 µW, and with a full-width-at-half-maximum beam diameter of ~4.3 µm. The duration of the laser pulse was limited by our instrumentation and can be reduced in the future using an optical modulator. The incident light induced the biased VO$_2$ device to transition to oscillation (O state) (denoted by Point 2 in Fig. 2A) with a DC electrical resistance reduction of $\Delta R$ = 28.9 kΩ (~31.2 % of the full transition). Figure 2B shows the DC voltage measured without (red) and with (blue) an incident laser pulse. In the blue curve, even after the incident writing laser was turned off, the original voltage in the $M_1$ phase was not recovered, and the device DC voltage remained at ~6.9 V for at least 1 hour (fig. S3). As we will discuss in the next paragraph, this DC voltage was the average of an oscillatory signal. Next, we removed $I_B$ for an arbitrarily long period of time, and we abruptly re-applied the current value within the 1st hysteresis, $I_B$, to read out the device state. We measured a DC voltage of ~6.9 V during the read time even after 3 days and without any other optical or electrical input. Figure 2C plots the first 30 min of the voltage readings taken every minute, showing the memory is nonvolatile. When the writing optical pulse was not applied, the device remained in the insulating phase with a voltage of ~9 V (red dots in Fig. 2C). When we repeated the experiment at room temperature by applying a bias current in the hysteresis of the 2nd transition (Fig. 2D) or a voltage bias (fig. S4 and Supplementary Text), the device completed the SPT and served as a volatile memory (i.e., the R phase could not persist without the current or voltage bias).

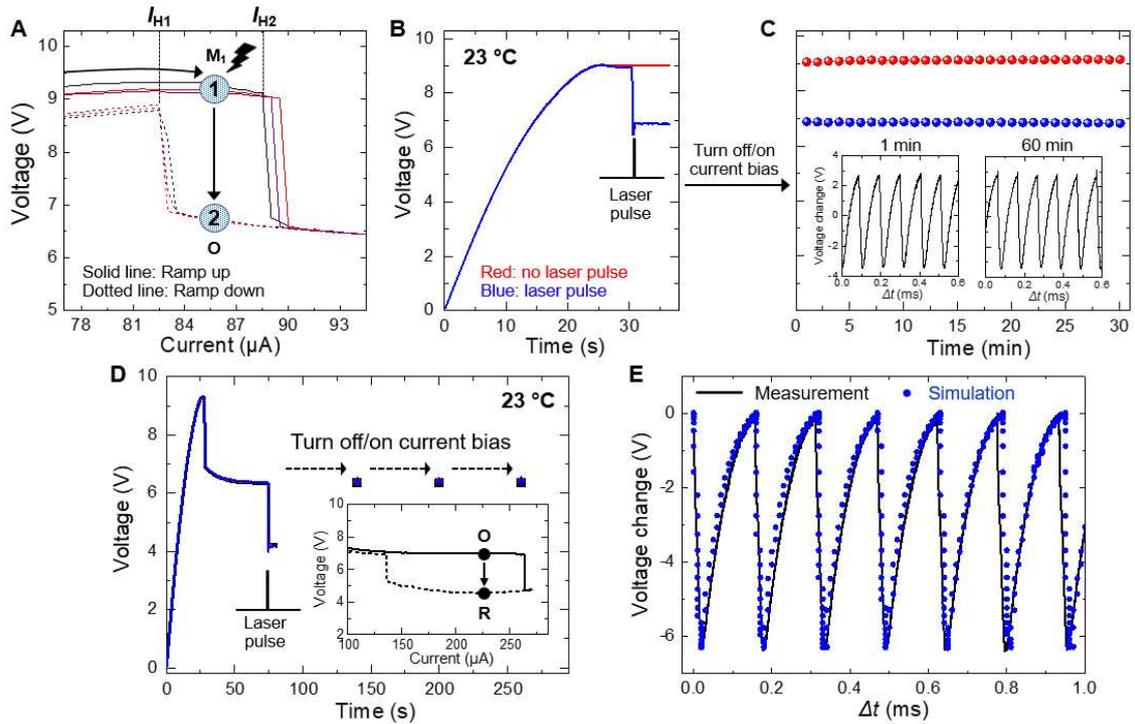

**Fig. 2. Operation of the VO₂ nonvolatile memory oscillator at a temperature of 23 °C.** (**A**) 1st hysteresis loop of the device (solid line: current ramp up, dashed line: current ramp down). The forward and backward transition currents, $I_{H1}$ and $I_{H2}$, are labelled. (**B**) Time trace of the DC voltage across the device during a ramp up of the bias current and holding at ~85 μA followed by no incident optical pulse (red), and a ~180 ms-long optical pulse at an incident optical power of ~470 μW (blue) showing optical memory writing. (**C**) The DC voltage readout after turning off and on the current bias every 1 minute. If an optical pulse was incident (blue dots), a voltage of ~6.7 V was read. If there was no light (red dots), the voltage of ~9.1 V was read. Inset, voltage oscillations corresponding to the DC voltage readout of the blue dots measured after 1 min and 60 min. (**D**) 3 overlapping real-time measurements of the device voltage using a bias current at the 2$^{nd}$ hysteresis (ramp up to ~230 μA) and an incident optical pulse with power of ~1.1 mW, including turning off and on the current bias. Inset, 2$^{nd}$ hysteresis loop of the device. The device returned to the O state indicated in the inset after the bias is removed. (**E**) Simulated and measured voltage traces in the O state. The trace is not centered at 0V since it was captured immediately after the transition and the DC component was not yet filtered.

Figure 2E shows the oscilloscope time trace of the sustained voltage oscillations in the O state, across our VO$_2$ device after the device transitioned to the intermediate state and the writing incident light was turned off. The oscillations with $I_B$ had a frequency of ~6.5 kHz and an amplitude of ~7 V, corresponding to the full voltage drop between the M$_1$ and R phases in Fig. 1D. The oscillation frequency in the absence of incident light was independent of the optical power of the write pulse but depended on the applied current bias. In addition, as shown in the inset of Fig. 2C, oscillations were repeatably observed each time when $I_B$ was applied for the voltage readout. Using a circuit model for our VO$_2$ device (fig. S5), we found that the simulated oscillations (blue dots in Fig. 2E) agree well with the measurements. The sustained oscillation frequency could be precisely tuned by the incident optical power of a continuous-wave laser after the memory writing process (fig. S6) and was dependent on the external circuit component, such as capacitor and resistor (*12*).

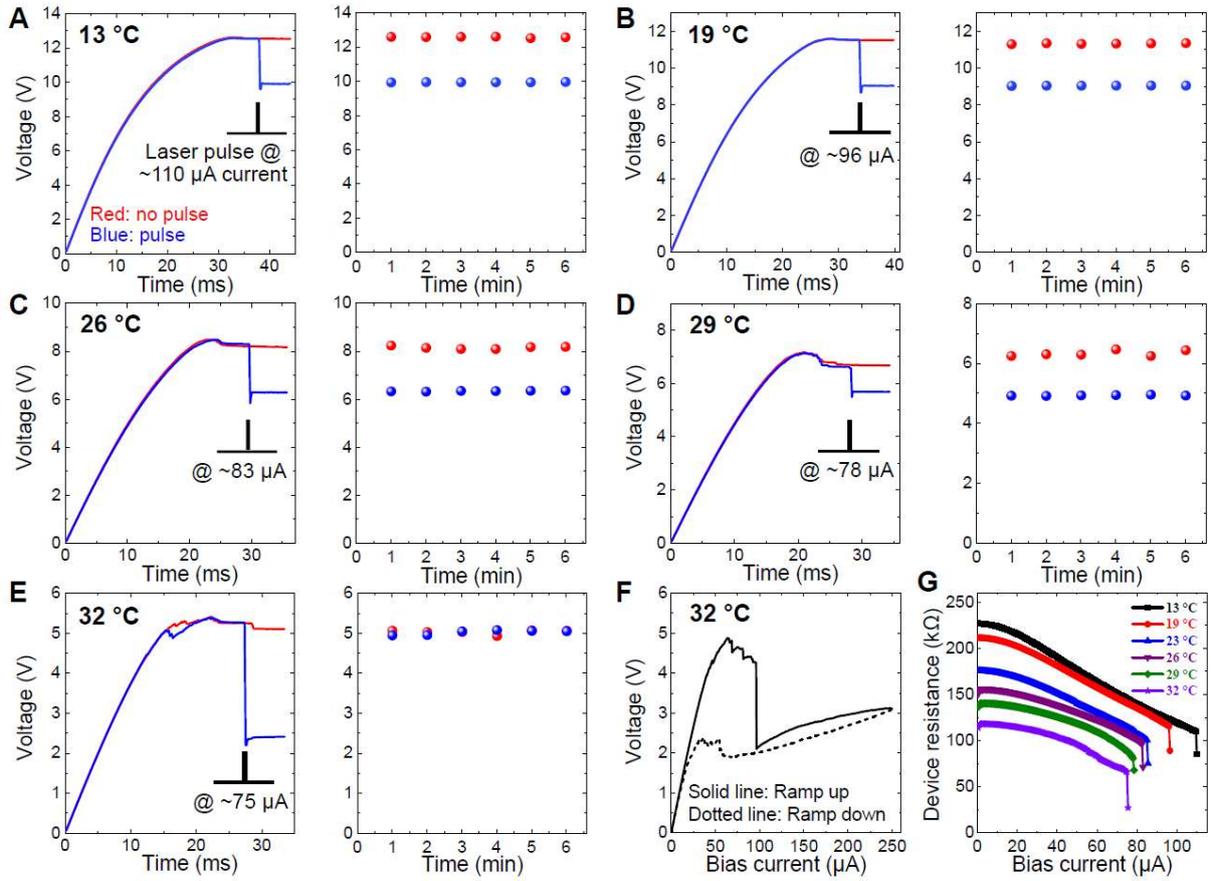

**Fig. 3. Temperature-dependent memory operation.** (**A-E**) At temperatures ranging from 13 °C to 32 °C, real-time voltage trace of the current-biased device with optical memory writing, and the tests of nonvolatile and volatile memory with the current bias turned off and on every minute. The identical optical pulse duration (~180 ms) and power (~470 μW) were used in the measurement at 13~29 °C. An incident laser power of ~1.1 mW was used at 32 °C for SPT. (**F**) VI measurement at the temperature of 32 °C. (**G**) Device resistance changes with bias currents and a light pulse at different temperatures.

To test the hypothesis that the nonvolatile memory would not be present if the thermal energy was sufficiently high to enable the full insulator-metal phase transition including the SPT (*24*), we repeated the experiment at different temperatures and bias conditions. As the sample stage temperature was increased from 13 °C to 29 °C (Fig. 3A-D), the current bias for the phase transition decreased due to the increase in thermally generated carriers (*25*), but the device could still function as a nonvolatile memory. When the sample stage was above 32 °C, the device could no longer function as a nonvolatile memory (Fig. 3E), and the SPT was evident from the VI characteristic (Fig. 3F). The device resistance as a function of the bias current and after an incident light pulse is shown in Fig. 3G. For stage temperatures between 13 °C and 29 °C, the insulating $VO_2$ (with device resistance, $R$, > 100 kΩ) transitioned to the O state ($R$ = 65 kΩ-85 kΩ) with the application of bias current and an optical pulse. At 32 °C, however, the device fully transitioned to the R phase (final $R$~25 kΩ). These observations are further supported by thermal simulations (fig. S7), which show that for a substrate temperature < ~30 °C, the maximum temperature reached in the device, even with an applied bias current for the first phase transition, remained below the SPT temperature.

**Threshold Optical Power**

A minimum optical power was found to be necessary to write to PEMO and initiate the oscillations. Figure 4A plots the probability of the phase transition vs. the peak power of a ~180 ms-long incident light pulse with the device biased at ~86 µA. The probability was obtained from 50 measurements at each optical power (fig. S8). The phase transition probability increased with the incident optical power due to the increased photo-induced carrier generation from the Mott transition. A threshold optical power of about 470 µW was required for reliable writing of the PEMO. The inset and right side of Fig. 4B show the voltage trace within the blue and red box on a magnified scale, respectively, and the sustained voltage oscillations were observed. The minimum optical power required for memory-writing can reduced by applying a bias current closer to the transition edge.

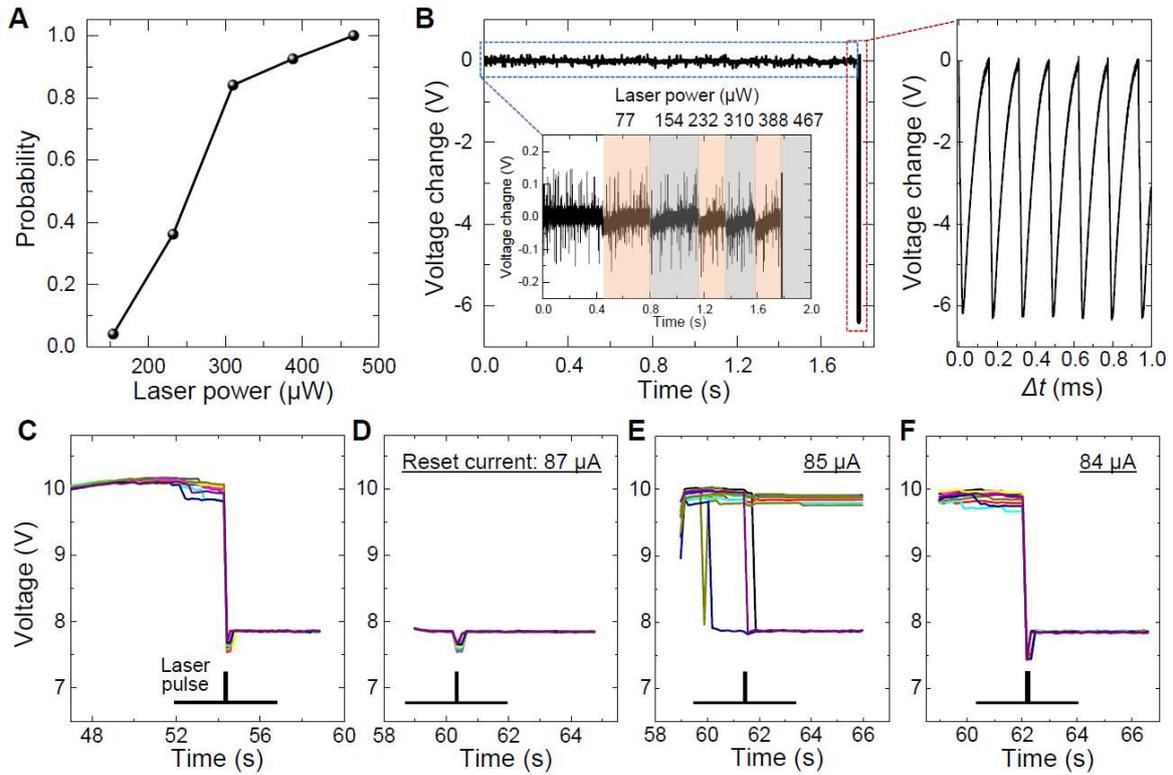

**Fig. 4. Optical threshold for the phase transition and memory reset test at room temperature**. (**A**) Probability that a ~180 ms-long light pulse at a wavelength of 1550nm induces the Mott transition at different optical powers. The results were determined from DC voltage measurements. (**B**) Voltage change across the device with incident laser pulses with varying powers. The magnified plots of blue and red box regions are in the inset and right side of (**B**), respectively. (**C**) Optical memory writing on a current-biased device at ~88 µA. (**D-F**) Erasing and rewriting test with different reset bias currents. 10 measurements are overlapped.

**Rise/Fall-Times for Memory Erasure and Readout**

To erase the PEMO, the bias current should be reduced to a value outside of the hysteresis. After writing to the PEMO using a laser pulse (Fig. 4C), we erased and rewrote to the device by applying reset currents and re-applying the ~180 ms-long and ~470 µW incident laser pulse (Fig. 4D-F). In this measurement, the device was initially biased at ~88 µA, since the hysteresis loop had slightly shifted to the range between ~85 µA and ~89 µA. When the device current was set to ~87 µA after the electrical input had been completely removed, the $VO_2$ was not erased because the reset current was inside the hysteresis (Fig. 4D). For a reset current of ~85 µA, which was closer to edge of the hysteresis, the device transitioned back to the insulating phase, but the optical rewriting after the bias current was re-applied was not always repeatable (Fig. 4E). For a lower reset current of ~84 µA outside the hysteresis, the reset and rewriting processes were repeatable (Fig. 4F). Finally, we characterized how fast the bias current must be turned on and off to read the memory without erasing it. By applying trapezoidal waveforms with different slopes, we found that the maximum the 0-to-100% rise/fall-time for the memory readout was about 150 µs (fig. S9).

**DISCUSSION AND CONCLUSION**

Our observation of nonvolatile memory suggests the presence of an intermediate metastable phase (M*) in $VO_2$ in the Mott transition. Without M*, when the read current is reapplied, the $VO_2$ should have returned to $M_1$ and no oscillations would be read out. We are currently investigating the structural and dynamic properties as well as the methods of control of M*. Nevertheless, the observations here show that M* can be leveraged to create nonvolatile memory devices. The memory can likely also be written using weak, short current pulses instead of light, since the writing depended on the generation of carriers rather than SPT. In contrast to $VO_2$ memristors based on SPT that would generally have thermally limited switching speed, the proposed PEMO dominantly utilizes an electronic phase (Mott) transition which can be addressed much faster than our demonstration, potentially less than ~1 ps (*26, 27*).

The recent years have witnessed a growing number of demonstrations of $VO_2$ memristors and $VO_2$ oscillator networks for neuromorphic computing (*5, 8-10*). $VO_2$ memristor-based artificial neurons have exhibited integrate-and-fire behaviors (*28*), as well as all three classes of excitability and most of the known biological neuronal dynamics (*5*). Using the phase differences between oscillations, image recognition has been performed by a three-oscillator-$VO_2$-network (*10*). These results demonstrate the exciting computing capabilities of small networks of $VO_2$ coupled oscillators, though without the capability of in-device memory. The observation of nonvolatile memory reported here can enable in-device memory in $VO_2$ to simplify system architectures and open new opportunities for phase-change photonics and electronics.

**Acknowledgments:** We thank A. Sharma and Y. Yang for help with device measurements.

**Funding:** The external funding sources were Natural Sciences and Engineering Research Council of Canada and the Canada Research Chair program.

# Supplementary Materials

Device fabrication

A 360 nm thick $VO_2$ film was formed on a titanium dioxide ((001) $TiO_2$) substrate by pulsed laser deposition (PLD) using a $VO_2$ target at a temperature of 380 °C. The devices were fabricated using aligned electron-beam lithography (EBL), dry etching, and metal deposition. First, 20 μm × 20 μm tungsten alignment markers for the EBL were sputtered onto the $VO_2$ film. Microwires, each of which is connected to rectangular areas of size 3 μm × 48 μm at the two ends forming 'H' shaped regions, were written by EBL (Raith EBPG 5000) using ZEP-520A positive tone resist and then inductively coupled plasma-reactive ion etching (ICP-RIE) was carried out. The rectangle/square at the two ends of the wires provides a large surface area between the $VO_2$ and the contact pads. A gas mixture of $CF_4$ (10 sccm) and Ar (20 sccm) at a total pressure of 50 mTorr was used. For a fully etched $VO_2$, an ICP power of 300 W and RF power of 50 W was applied for 30 s. Microwires with varying lengths of 2~5 μm and widths of 1~3 μm were fabricated. Finally, palladium (Pd) electrical contact pads were formed atop of the $VO_2$ through EBL, thermal evaporation deposition, and lift-off processes. Pd was chosen for electrical contacts with the $VO_2$ due to the low work function mismatch (*1*).

Measurement setup

The experimental setup is illustrated in Fig. S1. The sample was mounted on a temperature-controlled stage and accurately adjusted by a thermoelectric cooler (TEC) controller (5235 TEC Source, Arroyo Instruments) to 0.01 °C. The memory tests were performed in ambient conditions at temperatures ranging between 13 °C and 32 °C. The sourcing and measuring of device current/voltage were accomplished by contacting tungsten probes to the two Pd pads using a precision sourcemeter (2636A SourceMeter, Keithley) with a $R_{series}$=5 kΩ resistor in series, which protected the device from being overdriven during the phase transition. For the optical addressing of the device, a 1550-nm-wavelength optical beam from a tunable laser source (81600B, Agilent) was focused on the $VO_2$ wire through an 100X objective lens. The full-width at half maximum (FWHM) beam spot diameter was approximately 4.3 µm and the incident laser power on the device ranged from 80 µW to 2.5 mW. The incident laser pulses were approximately ~180 ms-long, limited by the time to turn the laser on and off via SCPI commands. A silicon CCD visible and InGaAs infrared cameras were used to align the device and laser beam spot, respectively. The generated voltage waveforms were measured by a 1 GHz bandwidth oscilloscope (InfiniiVision DSOX3104A, Keysight) connected to a $C_{ext}$=10 nF capacitor in parallel with the $VO_2$ device including $R_{series}$ via a 6 MHz oscilloscope probe (Tektronix 2220, 1× mode) (Fig. S1B). In Fig. 2E and Fig. 4B, the DC offset is not removed due to RC transients.

Volatile memory operation in a voltage-biased mode

Figure S4 shows the volatile memory using a constant-voltage bias at room temperature (23 °C). The overall operation of the voltage-biased device is similar to that of the current-biased one described in the main manuscript except a voltage is applied instead of a current. A DC current-voltage (IV) plot for a device with $VO_2$ wire dimensions of $w$ = ~1.7 µm and $l$ = ~1.7 µm was measured at room temperature (Fig. S4A), showing the full structural phase transition (SPT) to the metallic-rutile (R) phase at ~7.03 V and a larger hysteresis due to the predominantly thermally driven nature of this transition. For memory writing, we held the $VO_2$ wire at a voltage bias of ~6.6 V (denoted by Point 1 in Fig. S4A) and then illuminated with a laser pulse with an incident power of ~470 µW for inducing the transition (denoted by Point 2 in Fig. S4A). Fig. S4B plots time-dependent traces of the current changes of the device without (red) and with (blue) an incident laser pulse. After the laser was turned off, the original insulator-phase current was not recovered, and the device remained in the R phase with a measured DC current of ~1.13 mA. When different voltage biases within the hysteresis were applied (with accordingly adjusted incident laser power), memory writing was realized with the corresponding transition current levels (red dots in Fig S4A, average values of 50 measurements). To check whether the memory was volatile or nonvolatile, we removed the voltage bias and re-applied the same bias voltage, and the current reading reverted to the insulator-monoclinic (M1) phase current as shown in Fig. S4C. Thus, the memory was volatile. In this case, Joule heating mainly caused the SPT and the device was volatile because the R phase could not be sustained once the external stimuli (i.e., the voltage) was removed and the VO₂ temperature dropped to below the critical temperature.

Theoretical calculation of minimum optical power for insulator-metal transition (IMT)

To model the optical power required to initiate the IMT under a certain voltage bias, relationships between the external stimuli and VO₂ IMT conditions must be established first. According to Mott criterion, the carrier concentration dictates the VO₂ IMT as follows (*2, 3*):

$$n_c^{1/3} a_H = 0.25, \text{ where } a_H = 4\pi \frac{\varepsilon_0 \varepsilon_{VO_2} \hbar^2}{m_{VO_2} e^2}. \qquad \text{(Eq. S1)}$$

In Eq. S1, $n_c$ and $a_H$ stand for the critical carrier concentration and the Bohr radius of VO₂ where $\varepsilon_0$, $\varepsilon_{VO2}$, $\hbar$, $m_{VO2}$, and $e$ are the vacuum permittivity, the VO₂ relative DC permittivity, the reduced Planck constant, the effective mass of charged carriers in VO₂, and the elementary charge, respectively. By converting the electrical bias and optical stimulus into carrier concentrations and comparing the values with the critical carrier concentration, the required optical power under specific voltage/current biases for IMT can be calculated.

When the device is electrically biased, thermal- and field-induced carrier generation can both lead to an increase of the carrier concentration, $n$. Field-induced carrier generation is due to an autoionization caused by Coulomb barrier lowering, which is analogous to the Poole-Frenkel effect. In this case, the carrier concentration is given by (4)

$$n = N_0 \exp\left(-\frac{W - \beta\sqrt{E}}{KT_{VO_2}}\right), \text{ where } \beta = \sqrt{\frac{e^3}{\pi\varepsilon_{VO_2}\varepsilon_0}}. \qquad \text{(Eq. S2)}$$

In Eq. S2, $N_0$ is a constant independent of the electric field and temperature; $W$ and $\beta$ are the conductivity activation energy and the Poole-Frenkel constant, respectively; $E$, $k$ and $T_{VO2}$ are the electric field, the Boltzmann constant, and the local VO$_2$ temperature, respectively. Among the above parameters, $N_0$, $W$, and $\varepsilon_{VO2}$ are unknown constants with estimated ranges from various literature (0.15 eV < $W$ < 0.5 eV, 36 < $\varepsilon_{VO2}$ < 100) (4-7). $E$ can be estimated from the fabricated VO$_2$ wire length, $l$. To eliminate the unknown constant $N_0$, we consider the case when the carrier concentration reaches its critical value only resulting from measured transition temperature of $T_C$~314 K. Eq. S2 can be re-written as

$$n_c = N_0 \exp\left(-\frac{W}{KT_C}\right). \qquad \text{(Eq. S3)}$$

The expression for $n$ can be derived by dividing Eq. S2 and Eq. S3, which results in

$$n = n_c \exp\left\{-\frac{W}{KT_C}\left[\frac{T_C}{T_{VO_2}}\left(1 - \frac{\beta\sqrt{E}}{W}\right) - 1\right]\right\}. \qquad \text{(Eq. S4)}$$

To estimate $T_{VO2}$, which is difficult to measure accurately and directly, according to Fourier's law of heat conduction, the system can be described as

$$\frac{dQ}{dt} = P_{Joule} - k_{eff}\left(T_{VO_2} - T_A\right), \qquad (Eq.\ S5)$$

where $Q$, $P_{Joule}$, $k_{eff}$, and $T_A$ are the total heat of the system, Joule heating power, effective thermal conductance, and environment temperature, respectively. At equilibrium, $dQ/dt = 0$ and Eq. S5 becomes

$$T_{VO_2} = \frac{P_{Joule}}{k_{eff}} + T_A. \qquad (Eq.\ S6)$$

Considering that VO$_2$ undergoes IMT with a voltage bias $V_c$, Eq. S6 can be transformed into

$$k_{eff} = \frac{V_C^2/R_C}{T_{VO_2\_C} - T_A}, \qquad (Eq.\ S7)$$

where $R_C$ and $T_{VO2\_C}$ are the measurable resistance and non-measurable VO$_2$ local temperature, respectively, when the $V_c$ is applied. To solve for $T_{VO2\_C}$, we assumed that $n=nc$, $E=Ec=Vc/l$, and $T_{VO2} = T_{VO2\_C} < T_C$ (the transition temperature does not need to be as high due to the existence of an electric field) at the IMT and Eq. S4 becomes

$$T_{VO_2\_C} = T_C\left(1 - \frac{\beta\sqrt{E_C}}{W}\right) = T_C\left(1 - \frac{\beta\sqrt{V_C/l}}{W}\right) \qquad (Eq.\ S8)$$

By substituting Eq. S8 back into Eq. S7, then to Eq. S6, and finally to Eq. S4, the $n$ can be finally expressed by known constants, measurable variables, and constants with reported ranges ($W$ and $\varepsilon_{VO2}$). In addition, we also considered the current density equation, $J=e \cdot \mu \cdot n \cdot E$ where $\mu$ is the carrier mobility of VO$_2$ with a range between 0.15 and 0.3 cm$^2$/V·s.

The $W$, $\varepsilon_{VO2}$, and $\mu$ are variables that depend on various factors, for example, VO$_2$ crystallinity and the fabrication process. This study utilizes a parameter sweep method to find their values within the reported range that best fit the measured current-voltage properties of our fabricated VO$_2$ devices. The final values found are $W$=0.5 eV, $\varepsilon_{VO2}$=70, and $\mu$=0.265 cm$^2$/V·s, which are in the ranges of previous reports (*4-7*). As a result, the system is expressed by Eq. S4 and S8 and two variables $n$ and $J$ (or $E$), so VO$_2$ carrier concentration and current can be solved at each given voltage bias.

Furthermore, to convert the optical stimulus to carrier concentration, both photo-injection and photo-thermal effects need to be taken into consideration. When the fraction of laser power ($P_{laser}$) is absorbed, $\alpha$, of which $x$ is the fraction contributing to carrier generation (the remainder contributes to thermal generation), and the internal quantum efficiency is $\eta$, the carrier concentration increase can be expressed as

$$\Delta n = G_e \cdot \tau = \frac{\eta \cdot x \cdot \alpha \cdot P_{laser}}{hf} \cdot \tau \cdot \frac{1}{wlh}, \qquad \text{(Eq. S9)}$$

where $G_e$, $\tau$, $h$, and $f$ stand for the electron-hole pair generation rate, VO$_2$ carrier lifetime, Planck's constant, and laser frequency, respectively. $w$, $l$, and $h$ are the width, length, and height of the VO$_2$ microwire, respectively. The absorption, $\alpha$, is obtained from an electromagnetic wave finite-difference time-domain (FDTD) simulation to be ~19.5%, and $\tau$ is assumed to be 1 µs, which is acceptable compared to known values for VO$_2$ (*8, 9*). From photocurrent measurements at low intensities (such that $x$ can be assumed to be negligible) and assuming no photoconductive gain, the internal quantum efficiency is estimated to be about $\eta = 0.6\%$. The remaining optical power, $(1-x) \cdot \alpha \cdot P_{laser}$ contributes to the photo-thermal effect and Eq. S6 becomes

$$T_{VO_2} = \frac{P_{Joule} + (1-x) \cdot \alpha \cdot P_{laser}}{k_{eff}} + T_A. \qquad \text{(Eq. S10)}$$

By considering the difference between the critical carrier concentration and electrical bias-induced carrier concentration, we can theoretically determine the required optical power for IMT. The calculated minimum optical power for IMT, assuming $x = 5\%$, is plotted in Fig. S4D (black line) and agree well with the experimental results.

Circuit model for the electrical oscillation

The schematic of the circuit model for the oscillation is shown in Fig. S5, which is inspired by the driving point equivalent model (*10*). The VO$_2$ wire can be modeled as a resistor with two different resistance values and a capacitor in parallel. The material state is controlled by the switch, *S*, and the experimentally measured resistance values of the M$_1$ and R phases were used. To simulate the hysteresis, a bi-threshold control scheme is introduced for *S*. The switch closes when the voltage across the capacitor, $C_0$, is 1 V, and opens when the voltage is 0 V. Variable *comp$_1$* holds the voltage value across the VO$_2$, and *comp$_2$* supports the device state and phase transition value. Before the IMT when the VO$_2$ is in the insulator state, *comp$_2$* holds the value of IMT critical voltage, $V_H$, and voltage source $B_1$ is 0 V. When a sufficiently large electrical bias is applied such that the voltage across VO$_2$ is greater than $V_H$ *(comp$_1$>comp$_2$)*, *comp$_2$* changes to the metal-insulator-transition (MIT) critical voltage, $V_L$, voltage source $B_1$ raises to 1 V, and charges $C_0$ up with a time constant $\tau=R_0 \times C_0$ (simulates the IMT time). When the voltage across $C_0$ reaches 1 V, *S* closes and shorts out $R_1$, and VO$_2$ resistance changes to $R_{met}$, implying the IMT has taken place. When the device is in the metal state and $V_{VO2} = V_{comp1} < V_{comp2} = V_L$, *comp$_2$* switches to $V_L$, $B_1$ drops to 0 V, and discharges $C_0$. The *S* opens and the VO$_2$ resistance returns to $R_{ins}$.

To achieve oscillation, we considered the case where a current bias, $I_{osc}$, is applied such that $I_{osc}R_{ins} = V_{osc,ins} > V_H$ when VO$_2$ is in the M1 phase, and $I_{osc}R_{met} = V_{osc,met} < V_L$ in the R phase. In other words, this $I_{osc}$ is sufficient to initiate IMT but insufficient to maintain the metallic state due to its low metallic state resistance. As a result, MIT takes place immediately after IMT, and continues to oscillate as long as $I_{osc}$ is applied. As the only active component in the circuit, the device capacitance, $C_{VO2}$, determines the transient response time and the oscillation frequency by the repetition of discharging and charging processes. The respective equations are:

$$V = (V_H - V_L) \cdot \exp\left(-\frac{t}{\tau_{discharge}}\right) + V_L \quad \text{for discharging} \tag{Eq. S11}$$

$$V = (V_H - V_L) \cdot \left[1 - \exp\left(-\frac{t}{\tau_{charge}}\right)\right] + V_L \quad \text{for charging} \tag{Eq. S12}$$

with the time constants $\tau_{discharge} = R_{met}C_{VO2}$ and $\tau_{charge} = R_{ins}C_{VO2}$. A proper $C_{VO2}$ value needs to satisfy $C_{VO2} = \tau_{discharge}/R_{met} = \tau_{charge}/R_{ins}$ where $\tau_{discharge}$ and $\tau_{charge}$ are measured. The simulation result is in Fig. 2E.

Oscillation frequency control with a continuous wave (CW) laser

After the memory writing process, the sustained electrical oscillation frequency can be tuned by the excitation of a continuous-wave (CW) laser, resulting in the photo-induced carrier generation and photo-thermal effect. In this experiment, we use a fixed current bias of 86 μA, a 5 kΩ resistor of a resistor connected in series, and 10 nF capacitor in parallel with the VO$_2$ device including the series resistor. Figure S6A represents the response of the oscillation frequency as a function of CW laser power, and the tuning sensitivity of oscillation frequency was measured as ~0.84 kHz/mW with a great linear fit. Figures S6B and S6C show the generated waveforms with a frequency of 10.06 kHz and 11.77 kHz obtained with the laser incident power of ~390 μW and ~2.47 mW, respectively.

Thermal simulations

Numerical simulations based on the finite element method (FEM, COMSOL Multiphysics) and FDTD (Lumerical) were performed to quantify the heat generation. The width and length of the VO$_2$ wire placed on the TiO$_2$ substrate were assumed to be $w$ = 1.7 μm and $l$ = 4.7 μm, respectively. Before the thermal simulation, the electromagnetic radiation absorption at a VO$_2$ microwire was obtained from FDTD method by considering the measured (n, k) = (3.31, 0.377) (Fig. S2B) for the $VO_2$ in insulating phase at a wavelength of 1550 nm, which was used in our experiments. For the simulations, we assumed that the laser beam was focused on the $VO_2$ wire surface in Gaussian form and used ~4.3 μm (FWHM) that was measured from the InGaAs infrared camera. Assuming 30.6% of the incident optical power was absorbed (a result from the electromagnetic FDTD simulations), we calculated the heat generation in the $VO_2$ microwire with respect to the insulating and intermediate states to estimate the temperature distribution produced by the current-based (Joule heating) and photo-thermal heating processes, as shown in Fig. S7A. The thermal conductivity values were taken from (*11*). We assumed the substrate temperature to be 23 °C as in our experiment, and the simulated temperature of $VO_2$ wire rose to a maximum of 38.6 °C and then decreased to 35.5 °C after the 1$^{st}$ transition (Mott transition) due to the negative electrothermal feedback. Figure S7B shows corresponding temperature profile in the M1 phase. When compared to the resistance vs. temperature (RT) measurement (Fig. 1C), this thermal simulation shows that the device should not reach a sufficiently high temperature to undergo the 2$^{nd}$ transition (SPT), in which the transition temperature is ~41.5 °C. We also calculated the temperature of the $VO_2$ wire as a function of the substrate temperature as shown in Fig. S7C. The current bias and laser pulse values were chosen to approximate the experimental conditions (Fig. 3). At the substrate temperature range from 13 °C to 29 °C, the maximum temperature reached is still below the SPT temperature. However, when the substrate temperature was set to 32 °C, the device reached a peak temperature of 44.1 °C, sufficient for thermally driven SPT. Thus, the thermal simulation is in agreement with the experimental observation that the nonvolatile memory was absent for substrate temperatures ≥ 32 °C.

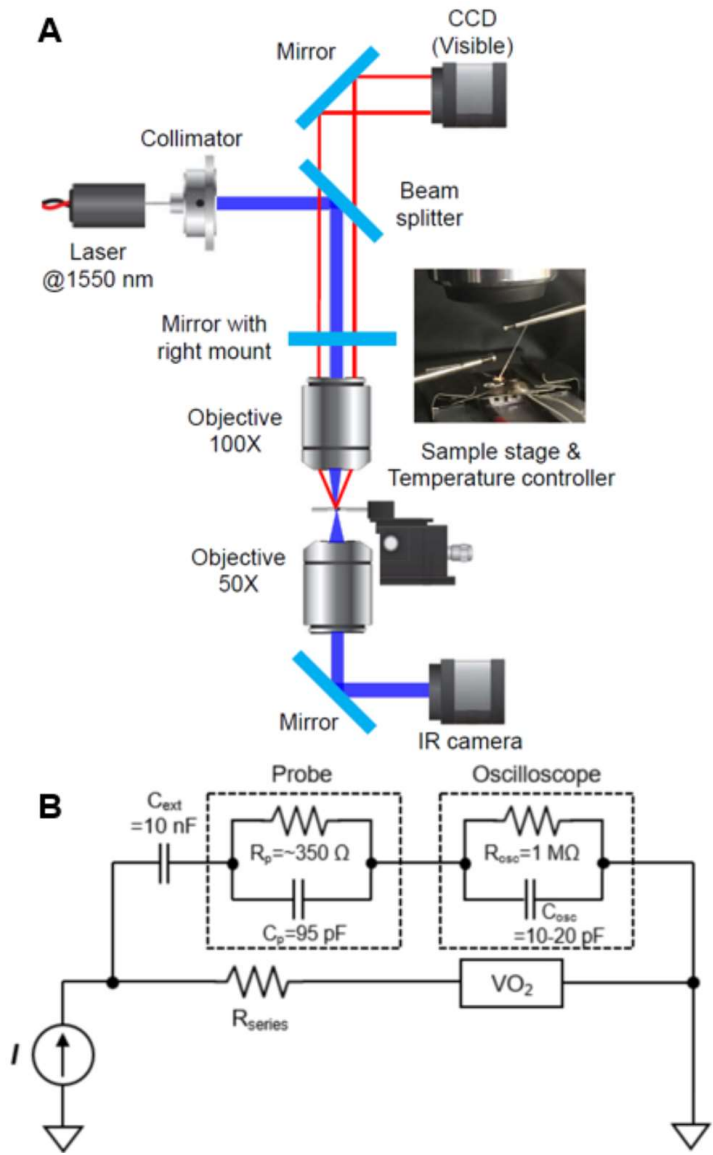

**Fig. S1. Experimental setup.** (**A**) Illustration for measuring the $VO_2$ PEMO device. (**B**) Circuit diagram of electrical measurement setup. The memory is written with an infrared laser with the device biased at a constant current.

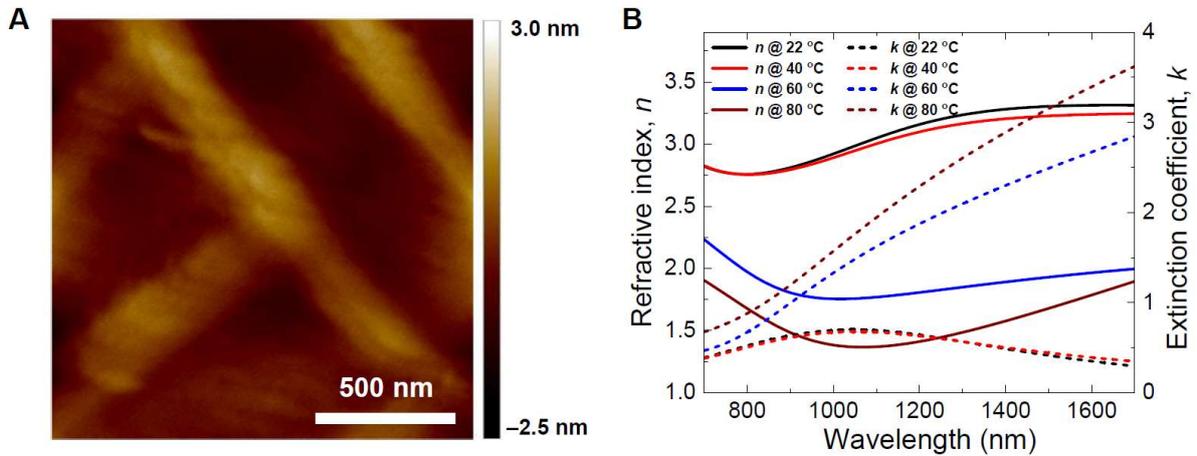

**Fig. S2. VO$_2$ film characterization.** (**A**) An atomic force microscope (AFM) tomography of the VO$_2$ film surface on TiO$_2$ substrate, which has a root-mean-square (RMS) roughness of ~0.55 nm. (**B**) Refractive index and extinction coefficient at various substrate temperature from ambient to ~80 °C. As the temperature increases, the refractive index decreases and the extinction coefficient strongly increases in the near infrared region, indicating the typical optical properties of VO$_2$ across its IMT.

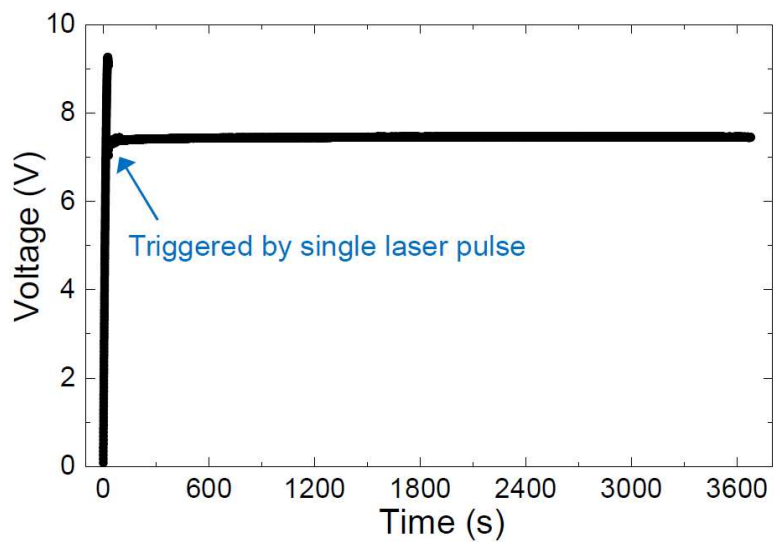

**Fig. S3. Real-time trace of the device DC voltage** after the optical memory writing. The device is maintained in the oscillating state for ~1 hour.

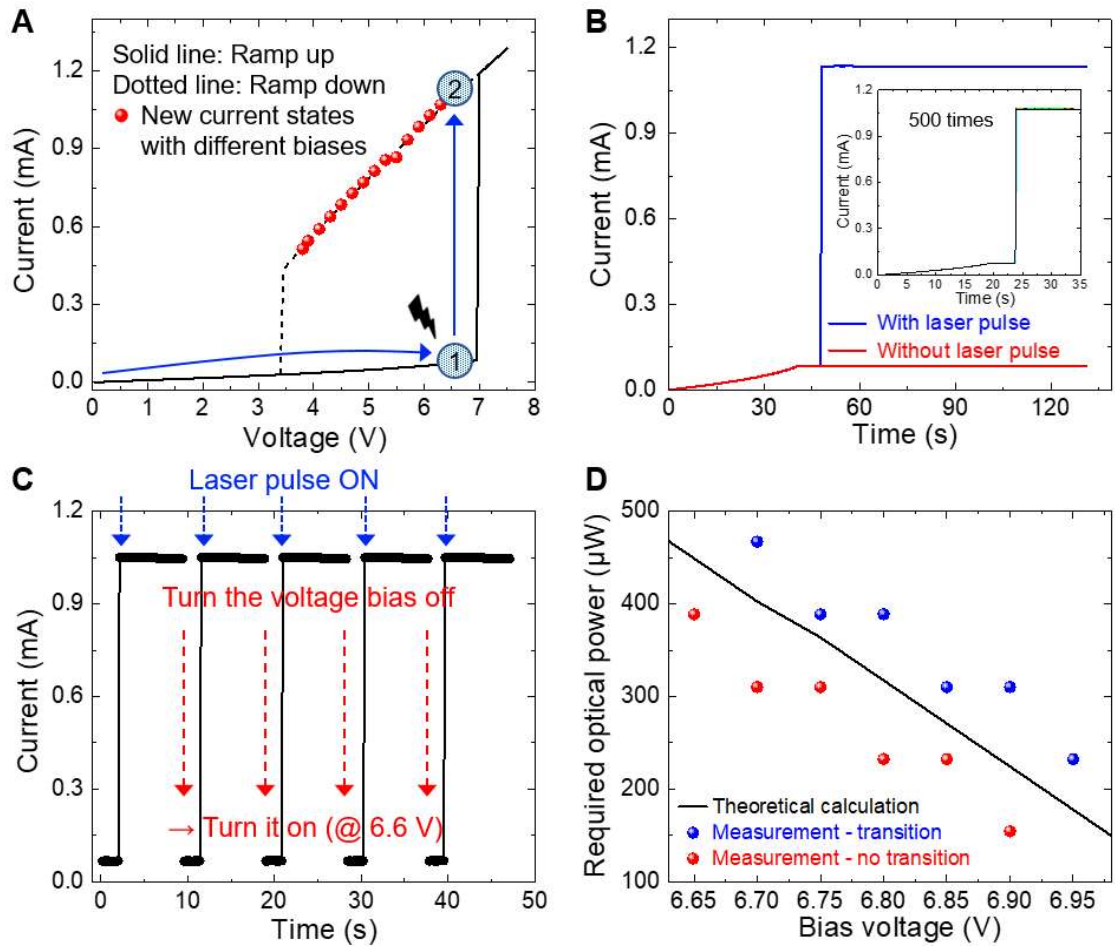

**Fig. S4. Volatile memory operation using a constant-voltage bias at a temperature of 23 °C.** (**A**) IV measurement (solid line: voltage ramp up, dashed line: voltage ramp down), showing an abrupt jump in current resulting from the full SPT. (**B**) Real-time trace of the device current, ramp up to ~6.6 V and then without laser pulse (red), and with a ~10 s long pulse at an incident optical power of ~470 μW (blue) showing optical memory writing. Inset, the overlap of 500 measurements. (**C**) Volatile switching operation. This memory writing and erasing were tested more than $10^4$ times without breakdown. (**D**) Theoretical calculation of the required optical power for memory writing as a function of the bias voltage along with experimental results.

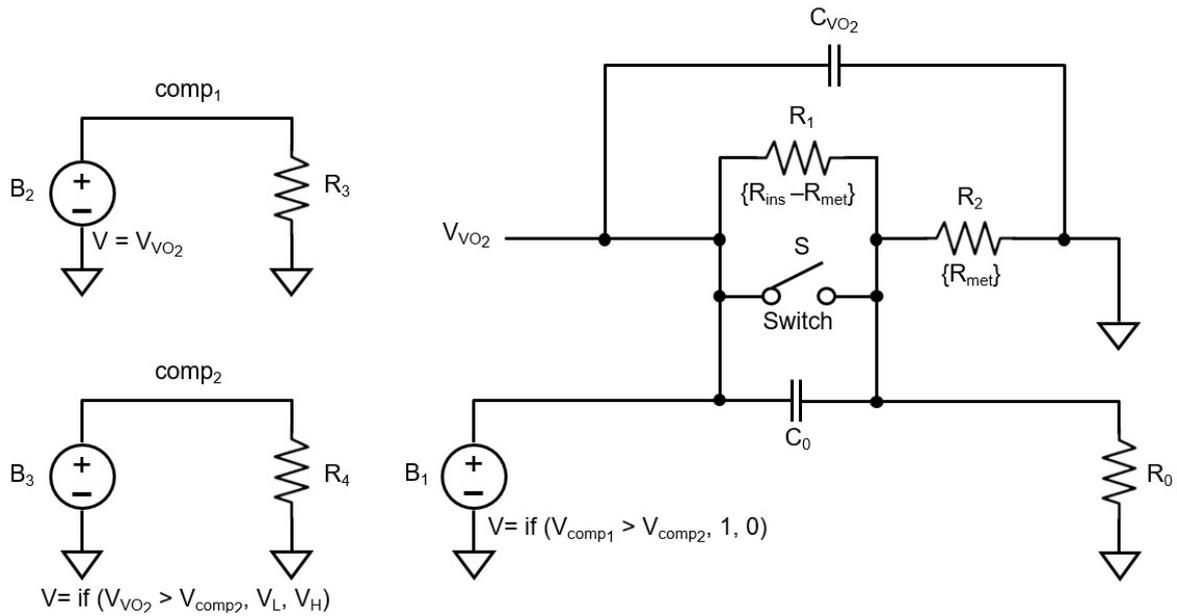

Fig. S5. Equivalent circuit model for the VO$_2$ microwire devices.

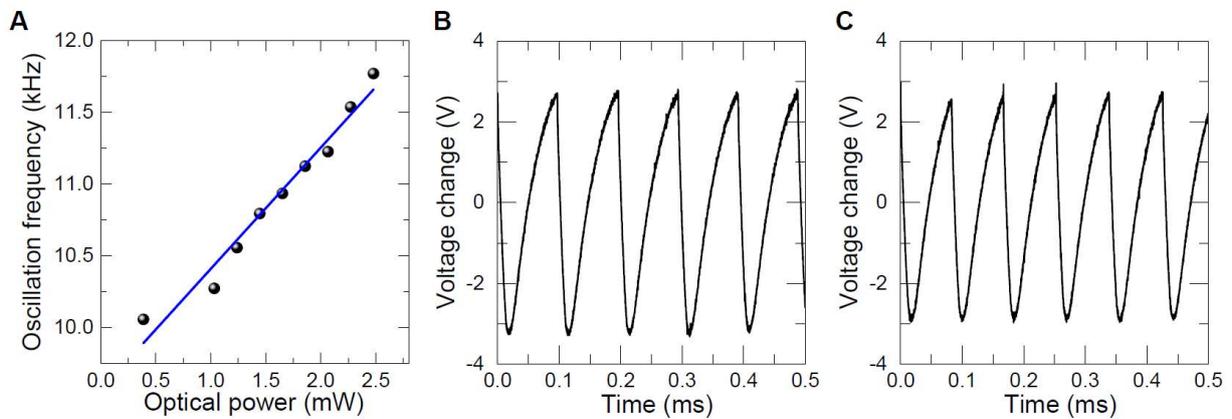

**Fig. S6. Oscillation frequency control by continuous wave laser beam power at room temperature. (A)** Measured electrical oscillation frequencies at different incident optical powers. The blue solid straight line represents a linear fit. **(B) and (C)**, Measured oscillations with an incident optical power of ~390 μW and ~2.47 mW, respectively. This measurement was performed using the same device described in the main manuscript.

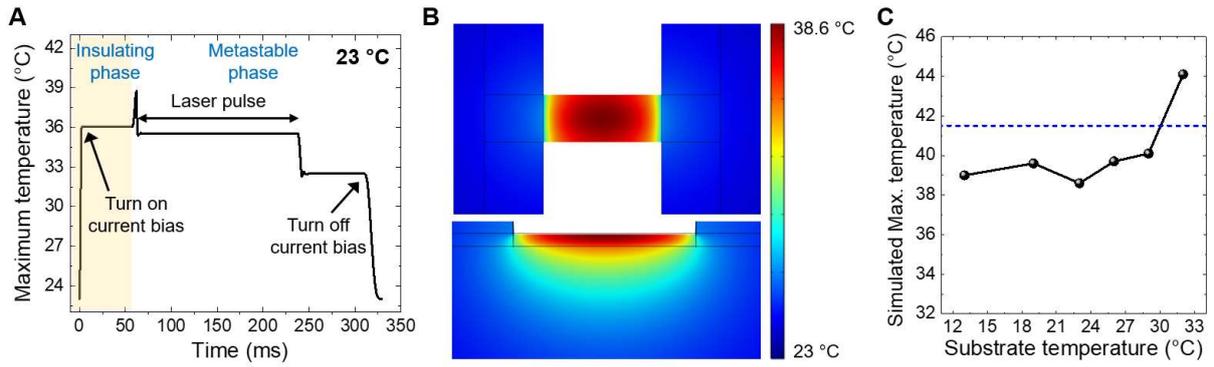

**Fig. S7. Thermal simulation.** (**A**) Time trace of the maximum temperature in VO$_2$ microwire by a current bias of ~85 μA and incident laser pulse power of 470 μW. (**B**) Simulated top (upper) and cross-sectional (bottom) temperature distribution for VO$_2$ wire of dimensions $w$=1.7 μm and $l$=4.7 μm as the laser pulse is applied on the M1 phase of VO$_2$. (**C**) Simulated maximum temperature of the VO$_2$ wire with varying the initial substrate temperatures. The dotted blue line indicates the phase transition temperature of IMT.

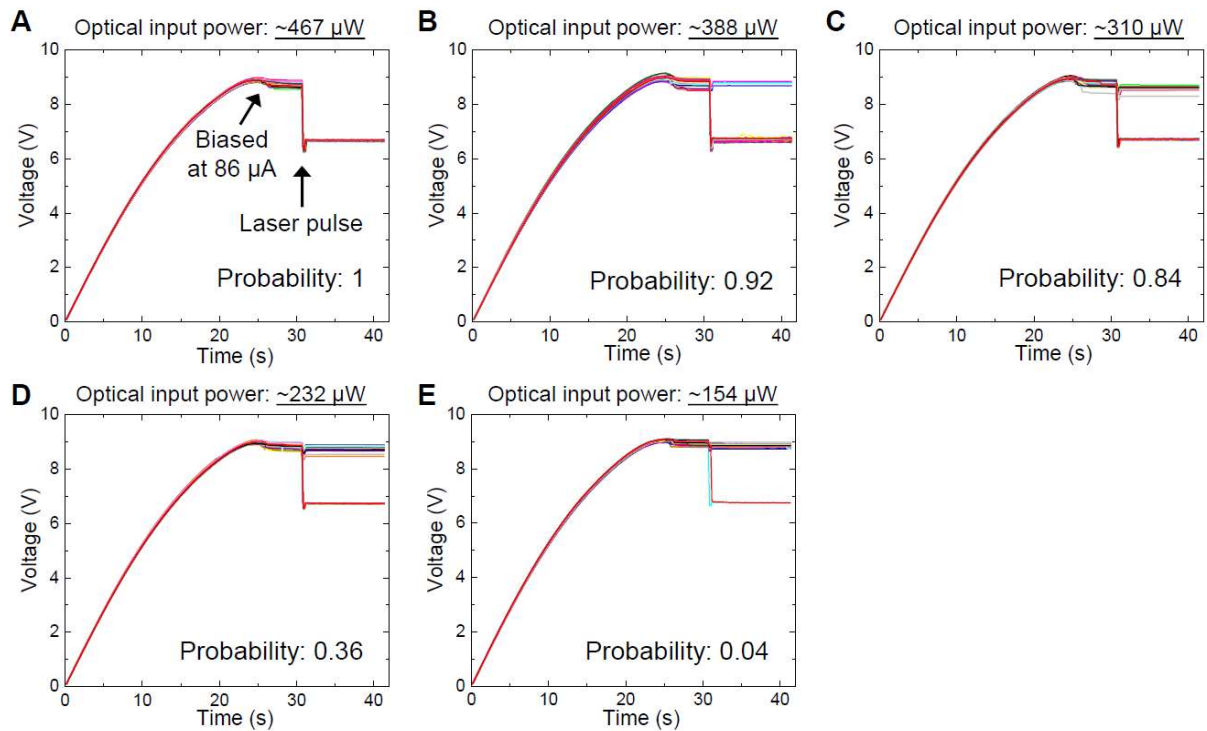

**Fig. S8. Optical threshold for the memory writing.** (A-E) Time trace of the device DC voltage with varying optical input power from ~470 µW to ~155 µW. The device was biased at 86 µA at room temperature (23 °C). 50 measurements were carried out at each optical power. The probability is significantly enhanced as the laser power is increased due to the escalated photo-induced carrier generation. At this current bias, the laser power of ~470 µW is necessary for the memory writing. If the electrical bias gets close to the 1st transition edge, a lower optical power is required.

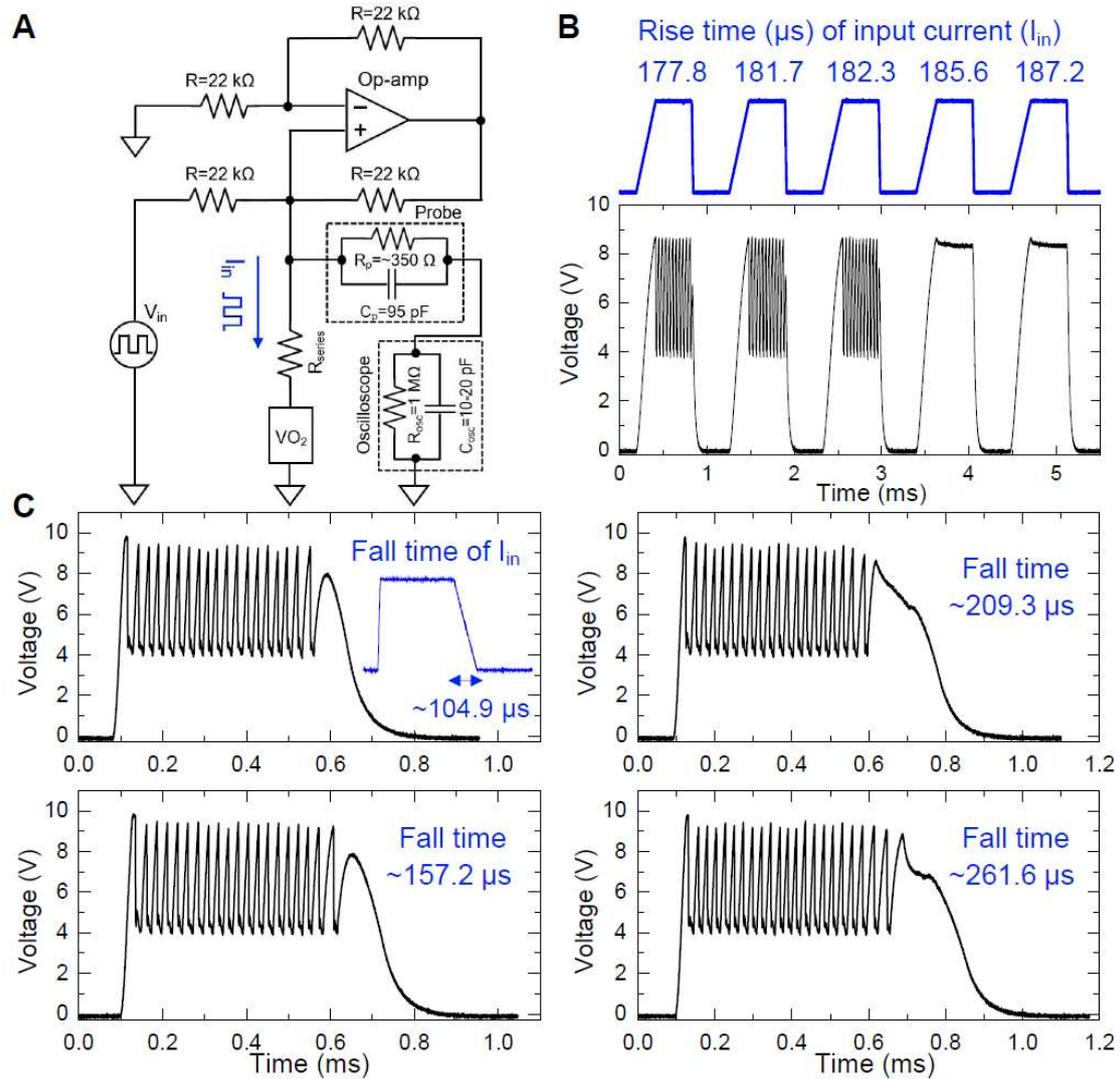

**Fig. S9. Rise/fall time of the applied current for reading and erasing the PEMO.** (**A**) Electrical circuit diagram for generating current pulses with varying rise/fall-times. After writing the memory so the VO$_2$ is in oscillation (O state), current pulses with varying rise/fall-times were applied to determine the maximum rise/fall-time that would reset the memory to the M$_1$ phase instead of preserving the oscillations. The amplitude of the current pulses was set to be equal to the original bias current. (**B**) and (**C**) show the real-time traces of the voltage across the device for varying rise and fall times, respectively. In (**B**), for 0%-to-100% rise-times greater than ~185.6 µs, voltage oscillations were no longer present, indicating that the device returned to the M$_1$. In (**C**), for 100%-

to-0% fall-times less than ~157 µs, the voltage reduced to 0 V in an exponential-like manner, regardless of when the current pulse was turned off during oscillation period. However, for 100%-to-0% fall-times longer than ~210 µs, the voltage decayed quasi-linearly before the exponential-like decay. The quasi-linear portion corresponds to the voltage adiabatically following the decrease in the current as in the static VI curve in Fig. 1D. Thus, (**B**) and (**C**) show that to preserve the nonvolatile memory, the PEMO should be read with current pulses with rise/fall-times less than about 150 µs.